\documentclass[12pt]{iopart}
\usepackage{graphicx}
\begin{document}
\title[Local Topological Constants for Non-Barotropic MHD]
{Variational Principles and Applications of Local Topological Constants of Motion for Non-Barotropic Magnetohydrodynamics}
\author{Asher Yahalom$^{a}$}
\address{Ariel University, Ariel 40700, Israel}
\ead{asya@ariel.ac.il}
\vspace{10pt}
\begin{indented}
\item[]December 2016
\end{indented}

%\maketitle

\newcommand{\beq} {\begin{equation}}
\newcommand{\enq} {\end{equation}}
\newcommand{\ber} {\begin {eqnarray}}
\newcommand{\enr} {\end {eqnarray}}
\newcommand{\eq} {equation}
\newcommand{\eqn} {equation }
\newcommand{\eqs} {equations }
\newcommand{\ens} {equations}
\newcommand{\mn}  {{\mu \nu}}
\newcommand {\er}[1] {equation (\ref{#1}) }
\newcommand {\eqref}[1] {equation (\ref{#1})}
\newcommand {\ern}[1] {equation (\ref{#1})}
\newcommand {\ers}[1] {equations (\ref{#1})}
\newcommand {\Er}[1] {Equation (\ref{#1}) }

\begin {abstract}

Variational principles for magnetohydrodynamics (MHD) were introduced by
previous authors both in Lagrangian and Eulerian form. In this
paper we introduce simpler Eulerian variational principles from
which all the relevant  equations of non-barotropic MHD can be derived for certain field topologies.
 The variational principle is given in terms of five independent functions
for non-stationary non-barotropic flows. This is less then the eight
variables which appear in the standard equations of barotropic
MHD which are the magnetic field $\vec B$ the
velocity field $\vec v$, the entropy $s$ and the density $\rho$.

The case of non-barotropic MHD in which the internal energy is a function of both entropy and density was not discussed in previous works which were concerned with the simplistic barotropic case. It is important to understand the rule of entropy and temperature for the variational analysis of MHD. Thus we introduce a variational principle of non-barotropic MHD and show that five functions will suffice to describe this physical system.

We will also discuss the implications of the above analysis for topological constants.
It will be shown that while cross helicity is not conserved for non-barotropic MHD a variant
of this quantity is. The implications of this to non-barotropic MHD stability is discussed.

\vspace{0.3cm}
\noindent Keywords: Magnetohydrodynamics, Variational principles, Topological conservation laws

\vspace{0.3cm}
\noindent PACS number(s): 47.65.+a
\end {abstract}

\section {Introduction}

Cross Helicity was first described by Woltjer \cite{Woltjer1,Woltjer2} and is give by:
\begin{equation} \label{GrindEQ__22_}
H_{C} \equiv \int  \vec{B}\cdot \vec{v}d^{3} x,
\end{equation}
in which $\vec{B}$ is the magnetic field, $\vec{v}$ is the velocity field and the integral is taken
over the entire flow domain. $H_{C}$ is conserved for barotropic or incompressible MHD and
is given a topological interpretation in terms of the knottiness of magnetic and flow field lines.
An analogous conserved helicity for fluid dynamics was obtained by Moffatt \cite{Moffatt}.
A generalization of barotropic fluid dynamics conserved quantities including helicity to non barotropic flows
including topological constants of motion is given by Mobbs \cite{Mobbs}. However, Mobbs did not
discuss the MHD case.

Both conservation laws for the helicity in the fluid dynamics case and the barotropic MHD case were
shown to originate from a relabelling symmetry through the Noether theorem \cite{Yahalomhel,Padhye1,Padhye2,YaLy}.
Webb et al. \cite{Webb2} have generalized the idea of relabelling symmetry to non-barotropic MHD and derived their generalized cross helicity conservation law by using Noether's theorem. The conservation law deduction involves a divergence symmetry of the action. These conservation laws were written as Eulerian conservation laws of the form $D_t+\vec \nabla \cdot \vec F = 0$ where D is the conserved density and F is the conserved flux. Webb et al. \cite{Webb4} discuss the cross helicity conservation law for non-barotropic MHD in a multi-symplectic formulation of MHD. Webb et al. \cite{Webb1,Webb2} emphasize that the generalized cross helicity conservation law, in MHD and the generalized helicity conservation law in non-barotropic fluids are non-local in the sense that they depend on the auxiliary nonlocal variable $\sigma$, which depends on the Lagrangian time integral of the temperature $T(x, t)$. Notice that a potential vorticity conservation equation for non-barotropic MHD is derived by Webb, G. M. and Mace, R.L. \cite{Webb5} by using Noether's second theorem.

It should be mentioned that Mobbs \cite{Mobbs} derived a helicity conservation law for ideal, non-barotropic fluid dynamics, which is of the same form as the cross helicity conservation law for non-barotropic MHD, except that the magnetic field induction is replaced by the generalized fluid helicity  $ \vec \Omega = \vec \nabla \times (\vec v - \sigma \vec \nabla s) $. Webb et al. \cite{Webb1,Webb2} also derive the Eulerian, differential form of Mobbs \cite{Mobbs} conservation law (although they did not reference Mobbs \cite{Mobbs}). Webb and Anco \cite{Webb3} show how Mobbs conservation law arises in multi-symplectic, Lagrangian fluid mechanics.

Variational principles for MHD were introduced by
previous authors both in Lagrangian and Eulerian form. Sturrock
\cite{Sturrock} has discussed in his book a Lagrangian variational
formalism for MHD. Vladimirov and Moffatt
\cite{Moffatt} in a series of papers have discussed an Eulerian
variational principle for incompressible MHD.
However, their variational principle contained three more
functions in addition to the seven variables which appear in the
standard equations of incompressible MHD which are the magnetic
field $\vec B$ the velocity field $\vec v$ and the pressure $P$.
Kats \cite{Kats} has generalized Moffatt's work for compressible
non barotropic flows but without reducing the number of functions
and the computational load. Moreover, Kats has shown that the
variables he suggested can be utilized to describe the motion of
arbitrary discontinuity surfaces \cite{Kats3,Kats4}. Sakurai
\cite{Sakurai} has introduced a two function Eulerian variational
principle for force-free MHD and used it as a
basis of a numerical scheme, his method is discussed in a book by
Sturrock \cite{Sturrock}. A method of solving the equations for
those two variables was introduced by Yang, Sturrock \& Antiochos
\cite{Yang}. Yahalom \& Lynden-Bell \cite{YaLy} combined the Lagrangian of
Sturrock \cite{Sturrock} with the Lagrangian of Sakurai
\cite{Sakurai} to obtain an {\bf Eulerian} Lagrangian principle for barotropic MHD
which will depend on only six functions. The variational
derivative of this Lagrangian produced all the equations
needed to describe barotropic MHD without any
additional constraints. The equations obtained resembled the
equations of Frenkel, Levich \& Stilman \cite{FLS} (see also \cite{Zakharov}).
Yahalom \cite{Yah} have shown that for the barotropic case four functions will
suffice. Moreover, it was shown that the cuts of some of those functions \cite{Yah2}
are topological local conserved quantities.

Variational principles of non
barotropic MHD can be found in the work of
Bekenstein \& Oron \cite{Bekenstien} in terms of 15 functions and
V.A. Kats \cite{Kats} in terms of 20 functions. The author of
this paper suspect that this number can be somewhat reduced.
Moreover, A. V. Kats  in a remarkable paper \cite{Kats2} (section
IV,E) has shown that there is a large symmetry group (gauge
freedom)  associated with the choice of those functions, this
implies that the number of degrees of freedom can be reduced.
Here we will show that only five functions will suffice to
describe non barotropic MHD in the case
that we enforce a Sakurai \cite{Sakurai} representation for the magnetic field.
Morrison \cite{Morrison} has suggested a Hamiltonian approach but this also depends on 8 canonical variables (see table 2 \cite{Morrison}).
In a series of papers Yahalom has suggest a five function variational principle for non-barotropic MHD \cite{Yahnbmhd,Yahalom2,Yahalom3,Yahalom4}

The plan of this paper is as follows: First we introduce the
standard notations and equations of non-barotropic
MHD. Next we introduce a generalization
of the barotropic variational principle suitable for the non-barotropic case.
Later we simplify the Eulerian variational principle and formulate it in terms of eight functions. Next we show
how three variational variables can be integrated algebraically thus reducing the variational principle
to five functions. Then we discuss the Aharanov-Bohm effect and analogous phenomena in non-barotropic MHD which are related
to the topological conservation laws of magnetic helicity and non-barotropic cross helicity. Finally we discuss the application of
those to the stability of MHD flows.

\section{Standard formulation of non-barotropic magnetohydrodynamics}

The standard set of \eqs solved for non-barotropic MHD are given below:
\beq
\frac{\partial{\vec B}}{\partial t} = \vec \nabla \times (\vec v \times \vec B),
\label{Beq}
\enq
\beq
\vec \nabla \cdot \vec B =0,
\label{Bcon}
\enq
\beq
\frac{\partial{\rho}}{\partial t} + \vec \nabla \cdot (\rho \vec v ) = 0,
\label{masscon}
\enq
\beq
\rho \frac{d \vec v}{d t}=
\rho (\frac{\partial \vec v}{\partial t}+(\vec v \cdot \vec \nabla)\vec v)  = -\vec \nabla p (\rho,s) +
\frac{(\vec \nabla \times \vec B) \times \vec B}{4 \pi}.
\label{Euler}
\enq
\beq
 \frac{d s}{d t}=0.
\label{Ent}
\enq
The following notations are utilized: $\frac{\partial}{\partial t}$ is the temporal derivative,
$\frac{d}{d t}$ is the temporal material derivative and $\vec \nabla$ has its
standard meaning in vector calculus. $\vec B$ is the magnetic field vector, $\vec v$ is the
velocity field vector, $\rho$ is the fluid density and $s$ is the specific entropy. Finally $p (\rho,s)$ is the pressure which
depends on the density and entropy (the non-barotropic case).

The justification for those \eqs and the conditions under which they apply can be found in standard books on MHD
(see for example \cite{Sturrock}).
The above applies to a collision-dominated plasma in local thermodynamic equilibrium.
Such conditions are seldom satisfied by physical plasmas, certainly not in astrophysics or in
 fusion-relevant magnetic confinement experiments. Never the less it is believed that the fastest
 macroscopic instabilities in those systems obey the above equations \cite{Yah2}, while instabilities
 associated with viscous or finite conductivity terms are slower. It should be noted that due to a
 theorem by Bateman \cite{Bateman} every physical system can be described by a variational principle (including viscous plasma) the
 trick is to find an elegant variational principle usually depending on a small amount of variational variables. The current
 work will discuss only ideal MHD while viscous MHD will be left for future endeavors.

 \Er{Beq}describes the
fact that the magnetic field lines are moving with the fluid elements ("frozen" magnetic field lines),
 \ern{Bcon} describes the fact that
the magnetic field is solenoidal, \ern{masscon} describes the conservation of mass and \ern{Euler}
is the Euler equation for a fluid in which both pressure
and Lorentz magnetic forces apply. The term:
\beq
\vec J =\frac{\vec \nabla \times \vec B}{4 \pi},
\label{J}
\enq
is the electric current density which is not connected to any mass flow.
\Er{Ent} describes the fact that heat is not created (zero viscosity, zero resistivity) in ideal non-barotropic MHD
and is not conducted, thus only convection occurs.
The number of independent variables for which one needs to solve is eight
($\vec v,\vec B,\rho,s$) and the number of \eqs (\ref{Beq},\ref{masscon},\ref{Euler},\ref{Ent}) is also eight.
Notice that \ern{Bcon} is a condition on the initial $\vec B$ field and is satisfied automatically for
any other time due to \ern{Beq}.

\section{Variational principle of non-barotropic magnetohydrodynamics}

In the following section we will generalize the approach of \cite{YaLy} for the non-barotropic case.
Consider the action:
\ber A & \equiv & \int {\cal L} d^3 x dt,
\nonumber \\
{\cal L} & \equiv & {\cal L}_1 + {\cal L}_2,
\nonumber \\
{\cal L}_1 & \equiv & \rho (\frac{1}{2} \vec v^2 - \varepsilon (\rho,s)) +  \frac{\vec B^2}{8 \pi},
\nonumber \\
{\cal L}_2 & \equiv & \nu [\frac{\partial{\rho}}{\partial t} + \vec \nabla \cdot (\rho \vec v )]
- \rho \alpha \frac{d \chi}{dt} - \rho \beta \frac{d \eta}{dt} - \rho \sigma \frac{d s}{dt}
 - \frac{\vec B}{4 \pi} \cdot \vec \nabla \chi \times \vec \nabla \eta.
\label{Lagactionsimp}
\enr
In the above $\varepsilon$ is the specific internal energy (internal energy per unit of mass).
The reader is reminded of the following thermodynamic relations which
will become useful later:
\ber
d \varepsilon &=& T ds - P d \frac{1}{\rho} = T ds + \frac{P}{\rho^2} d \rho
\nonumber \\
& & \frac{\partial \varepsilon}{\partial s} = T, \qquad \frac{\partial \varepsilon}{\partial \rho} = \frac{P}{\rho^2}
\nonumber \\
w &=& \varepsilon + \frac{P}{\rho}= \varepsilon + \frac{\partial \varepsilon}{\partial \rho} \rho = \frac{\partial (\rho \varepsilon)}{\partial \rho}
\nonumber \\
dw &=& d\varepsilon + d(\frac{P}{\rho}) = T ds +  \frac{1}{\rho} dP
\label{thermodyn}
\enr
in the above $T$ is the temperature and $w$ is the specific enthalpy.
Obviously $\nu,\alpha,\beta,\sigma$ are Lagrange multipliers which were inserted in such a
way that the variational principle will yield the following \ens:
\ber
& & \frac{\partial{\rho}}{\partial t} + \vec \nabla \cdot (\rho \vec v ) = 0,
\nonumber \\
& & \rho \frac{d \chi}{dt} = 0,
\nonumber \\
& & \rho \frac{d \eta}{dt} = 0.
\nonumber \\
& & \rho \frac{d s}{dt} = 0.
\label{lagmul}
\enr
It {\bf is not} assumed that $\nu,\alpha,\beta,\sigma$  are single valued.
Provided $\rho$ is not null those are just the continuity \ern{masscon}, entropy conservation
 and the conditions that Sakurai's functions are comoving.
Taking the variational derivative with respect to $\vec B$ we see that
\beq
\vec B = \hat {\vec B} \equiv \vec \nabla \chi \times \vec \nabla \eta.
\label{Bsakurai2}
\enq
Hence $\vec B$ is in Sakurai's form and satisfies \ern{Bcon}.
It can be easily shown that provided that $\vec B$ is in the form given in \ern{Bsakurai2},
and \ers{lagmul} are satisfied, then also \ern{Beq} is satisfied.

For the time being we have showed that all the equations of non-barotropic MHD can be obtained
from the above variational principle except Euler's equations. We will now
show that Euler's equations can be derived from the above variational principle
as well. Let us take an arbitrary variational derivative of the above action with
respect to $\vec v$, this will result in:
\beq
\hspace{-2cm}\delta_{\vec v} A = \int dt \{ \int d^3 x dt \rho \delta \vec v \cdot
[\vec v - \vec \nabla \nu - \alpha \vec \nabla \chi - \beta \vec \nabla \eta - \sigma \vec \nabla s]
+ \oint d \vec S \cdot \delta \vec v \rho \nu+  \int d \vec \Sigma \cdot \delta \vec v \rho [\nu]\}.
\label{delActionv}
\enq
The integral $\oint d \vec S \cdot \delta \vec v \rho \nu$ vanishes in many physical scenarios.
In the case of astrophysical flows this integral will vanish since $\rho=0$ on the flow
boundary, in the case of a fluid contained
in a vessel no flux boundary conditions $\delta \vec v \cdot \hat n =0$ are induced
($\hat n$ is a unit vector normal to the boundary). The surface integral $\int d \vec \Sigma$
 on the cut of $\nu$ vanishes in the case that $\nu$ is single valued and $[\nu]=0$ as is the case for
 some flow topologies. In the case that $\nu$ is not single valued only a Kutta type velocity perturbation \cite{YahPinhasKop} in which
the velocity perturbation is parallel to the cut will cause the cut integral to vanish. An arbitrary
velocity perturbation on the cut will indicate that $\rho=0$ on this surface which is contradictory to
the fact that a cut surface is to some degree arbitrary as is the case for the zero line of an azimuthal
angle. We will show later that the "cut" surface is co-moving with the flow hence it may become quite complicated.
This uneasy situation may be somewhat be less restrictive when the flow has some symmetry properties.

Provided that the surface integrals do vanish and that $\delta_{\vec v} A =0$ for an arbitrary
velocity perturbation we see that $\vec v$ must have the following form:
\beq
\vec v = \hat {\vec v} \equiv \vec \nabla \nu + \alpha \vec \nabla \chi + \beta \vec \nabla \eta + \sigma \vec \nabla s.
\label{vform}
\enq
The above equation is reminiscent of Clebsch representation in non magnetic fluids \cite{Clebsch1,Clebsch2}.
Let us now take the variational derivative with respect to the density $\rho$ we obtain:
\ber
\delta_{\rho} A & = & \int d^3 x dt \delta \rho
[\frac{1}{2} \vec v^2 - w  - \frac{\partial{\nu}}{\partial t} -  \vec v \cdot \vec \nabla \nu]
\nonumber \\
 & + & \int dt \oint d \vec S \cdot \vec v \delta \rho  \nu +
  \int dt \int d \vec \Sigma \cdot \vec v \delta \rho  [\nu] + \int d^3 x \nu \delta \rho |^{t_1}_{t_0}.
\label{delActionrho}
\enr
In which $ w= \frac{\partial (\varepsilon \rho)}{\partial \rho}$ is the specific enthalpy.
Hence provided that $\oint d \vec S \cdot \vec v \delta \rho  \nu$ vanishes on the boundary of the domain
and $ \int d \vec \Sigma \cdot \vec v \delta \rho  [\nu]$ vanishes on the cut of $\nu$
in the case that $\nu$ is not single valued\footnote{Which entails either a Kutta type
condition for the velocity in contradiction to the "cut" being an arbitrary surface, or a vanishing density perturbation on the cut.}
and in initial and final times the following \eqn must be satisfied:
\beq
\frac{d \nu}{d t} = \frac{1}{2} \vec v^2 - w, \qquad
\label{nueq}
\enq
Since the right hand side of the above equation is single valued as it is made of physical quantities, we conclude that:
\beq
\frac{d [\nu]}{d t} = 0.
\label{mplicated nueqc}
\enq
Hence the cut value is co-moving with the flow and thus the cut surface may become arbitrary complicated.
This uneasy situation may be somewhat be less restrictive when the flow has some symmetry properties.

Finally we have to calculate the variation with respect to both $\chi$ and $\eta$
this will lead us to the following results:
\ber
\hspace{-2cm} \delta_{\chi} A  & \hspace{-1cm} = & \hspace{-0.8cm} \int d^3 x dt \delta \chi
[\frac{\partial{(\rho \alpha)}}{\partial t} +  \vec \nabla \cdot (\rho \alpha \vec v)-
\vec \nabla \eta \cdot \vec J]
+ \int dt \oint d \vec S \cdot [\frac{\vec B}{4 \pi} \times \vec \nabla \eta - \vec v \rho \alpha]\delta \chi
 \nonumber \\
 & + & \int dt \int d \vec \Sigma \cdot [\frac{\vec B}{4 \pi} \times \vec \nabla \eta - \vec v \rho \alpha][\delta \chi]
 - \int d^3 x \rho \alpha \delta \chi |^{t_1}_{t_0},
\label{delActionchi}
\enr
\ber
\hspace{-2cm} \delta_{\eta} A  &\hspace{-1cm} = & \hspace{-0.8cm} \int d^3 x dt \delta \eta
[\frac{\partial{(\rho \beta)}}{\partial t} +  \vec \nabla \cdot (\rho \beta \vec v)+
\vec \nabla \chi \cdot \vec J]
+ \int dt \oint d \vec S \cdot [\vec \nabla \chi \times \frac{\vec B}{4 \pi} - \vec v \rho \beta]\delta \eta
\nonumber \\
 & + &  \int dt \int d \vec \Sigma \cdot [\vec \nabla \chi \times \frac{\vec B}{4 \pi} - \vec v \rho \beta][\delta \eta]
 - \int d^3 x \rho \beta \delta \eta |^{t_1}_{t_0}.
\label{delActioneta}
\enr
Provided that the correct temporal and boundary conditions are met with
respect to the variations $\delta \chi$ and $\delta \eta$ on the domain boundary and
on the cuts in the case that some (or all) of the relevant functions are non single valued.
we obtain the following set of equations:
\beq
\frac{d \alpha}{dt} = \frac{\vec \nabla \eta \cdot \vec J}{\rho}, \qquad
\frac{d \beta}{dt} = -\frac{\vec \nabla \chi \cdot \vec J}{\rho},
\label{albetaeq}
\enq
in which the continuity \ern{masscon} was taken into account. By correct temporal conditions we
mean that both $\delta \eta$ and $\delta \chi$ vanish at initial and final times. As for boundary
conditions which are sufficient to make the boundary term vanish on can consider the case that
the boundary is at infinity and both $\vec B$ and $\rho$ vanish. Another possibility is that the boundary is
impermeable and perfectly conducting. A sufficient condition for the integral over the "cuts" to vanish
is to use variations $\delta \eta$ and $\delta \chi$ which are single valued. It can be shown that
$\chi$ can always be taken to be single valued, hence taking $\delta \chi$ to be single valued is no
restriction at all. In some topologies $\eta$ is not single valued and in those cases a single valued
restriction on $\delta \eta$ is sufficient to make the cut term null.

Finally we take a variational derivative with respect to the entropy $s$:
\ber
\delta_{s} A \hspace{-0.4cm} & = & \hspace{-0.4cm} \int d^3 x dt \delta s
[\frac{\partial{(\rho \sigma)}}{\partial t} +  \vec \nabla \cdot (\rho \sigma \vec v)- \rho T]
+ \int dt \oint d \vec S \cdot \rho \sigma \vec v  \delta s
 \nonumber \\
 & - &  \int d^3 x \rho \sigma \delta s |^{t_1}_{t_0},
\label{delActions}
\enr
in which the temperature is $T=\frac{\partial \varepsilon}{\partial s}$. We notice that according
to \ern{vform} $\sigma$ is single valued and hence no cuts are needed. Taking into account the continuity
\ern{masscon} we obtain for locations in which the density $\rho$ is not null the result:
\beq
\frac{d \sigma}{dt} =T,
\label{sigmaeq}
\enq
provided that $\delta_{s} A$ vanished for an arbitrary $\delta s$.

\section{Euler's equations}

We shall now show that a velocity field given by \ern{vform}, such that the
\eqs for $\alpha, \beta, \chi, \eta, \nu, \sigma, s$ satisfy the corresponding equations
(\ref{lagmul},\ref{nueq},\ref{albetaeq},\ref{sigmaeq}) must satisfy Euler's equations.
Let us calculate the material derivative of $\vec v$:
\beq
\frac{d\vec v}{dt} = \frac{d\vec \nabla \nu}{dt}  + \frac{d\alpha}{dt} \vec \nabla \chi +
 \alpha \frac{d\vec \nabla \chi}{dt}  +
\frac{d\beta}{dt} \vec \nabla \eta + \beta \frac{d\vec \nabla \eta}{dt}+\frac{d\sigma}{dt} \vec \nabla s +
\sigma \frac{d\vec \nabla s}{dt}.
\label{dvform}
\enq
It can be easily shown that:
\ber
\frac{d\vec \nabla \nu}{dt} & = & \vec \nabla \frac{d \nu}{dt}- \vec \nabla v_k \frac{\partial \nu}{\partial x_k}
 = \vec \nabla (\frac{1}{2} \vec v^2 - w)- \vec \nabla v_k \frac{\partial \nu}{\partial x_k},
 \nonumber \\
 \frac{d\vec \nabla \eta}{dt} & = & \vec \nabla \frac{d \eta}{dt}- \vec \nabla v_k \frac{\partial \eta}{\partial x_k}
 = - \vec \nabla v_k \frac{\partial \eta}{\partial x_k},
 \nonumber \\
 \frac{d\vec \nabla \chi}{dt} & = & \vec \nabla \frac{d \chi}{dt}- \vec \nabla v_k \frac{\partial \chi}{\partial x_k}
 = - \vec \nabla v_k \frac{\partial \chi}{\partial x_k},
  \nonumber \\
 \frac{d\vec \nabla s}{dt} & = & \vec \nabla \frac{d s}{dt}- \vec \nabla v_k \frac{\partial s}{\partial x_k}
 = - \vec \nabla v_k \frac{\partial s}{\partial x_k}.
 \label{dnabla}
\enr
In which $x_k$ is a Cartesian coordinate and a summation convention is assumed. Inserting the result from equations (\ref{dnabla},\ref{lagmul})
into \ern{dvform} yields:
\ber
\frac{d\vec v}{dt} &=& - \vec \nabla v_k (\frac{\partial \nu}{\partial x_k} + \alpha \frac{\partial \chi}{\partial x_k} +
\beta \frac{\partial \eta}{\partial x_k} + \sigma \frac{\partial s}{\partial x_k}) + \vec \nabla (\frac{1}{2} \vec v^2 - w)+ T \vec \nabla s
 \nonumber \\
&+& \frac{1}{\rho} ((\vec \nabla \eta \cdot \vec J)\vec \nabla \chi - (\vec \nabla \chi \cdot \vec J)\vec \nabla \eta)
 \nonumber \\
&=& - \vec \nabla v_k v_k + \vec \nabla (\frac{1}{2} \vec v^2 - w) + T \vec \nabla s
 + \frac{1}{\rho} \vec J \times (\vec \nabla \chi \times  \vec \nabla \eta)
 \nonumber \\
&=& - \frac{\vec \nabla p}{\rho} + \frac{1}{\rho} \vec J \times \vec B.
\label{dvform2}
\enr
In which we have used both \ern{vform} and \ern{Bsakurai2} in the above derivation. This of course
proves that the non-barotropic Euler equations can be derived from the action given in \er{Lagactionsimp} and hence
all the equations of non-barotropic MHD can be derived from the above action
without restricting the variations in any way except on the relevant boundaries and cuts.

\section{Simplified action}

The reader of this paper might argue here that the paper is misleading. The author has declared
that he is going to present a simplified action for non-barotropic MHD instead he
 added six more functions $\alpha,\beta,\chi,\-\eta,\nu,\sigma$ to the standard set $\vec B,\vec v,\rho,s$.
In the following I will show that this is not so and the action given in \ern{Lagactionsimp} in
a form suitable for a pedagogic presentation can indeed be simplified. It is easy to show
that the Lagrangian density appearing in \ern{Lagactionsimp} can be written in the form:
\ber
{\cal L} & = & -\rho [\frac{\partial{\nu}}{\partial t} + \alpha \frac{\partial{\chi}}{\partial t}
+ \beta \frac{\partial{\eta}}{\partial t}+ \sigma \frac{\partial{s}}{\partial t}+\varepsilon (\rho,s)] +
\frac{1}{2}\rho [(\vec v-\hat{\vec v})^2-(\hat{\vec v})^2]
\nonumber \\
& + &   \frac{1}{8 \pi} [(\vec B-\hat{\vec B})^2-(\hat{\vec B})^2]+
\frac{\partial{(\nu \rho)}}{\partial t} + \vec \nabla \cdot (\nu \rho \vec v ).
\label{Lagactionsimp4}
\enr
In which $\hat{\vec v}$ is a shorthand notation for $\vec \nabla \nu + \alpha \vec \nabla \chi +
 \beta \vec \nabla \eta +  \sigma \vec \nabla s $ (see \ern{vform}) and $\hat{\vec B}$ is a shorthand notation for
 $\vec \nabla \chi \times \vec \nabla \eta$ (see \ern{Bsakurai2}). Thus ${\cal L}$ has four contributions:
\ber
 \hspace{-2.2 cm}  {\cal L}  &  = &  \hat {\cal L} + {\cal L}_{\vec v}+ {\cal L}_{\vec B}+{\cal L}_{boundary},
\nonumber \\
\hspace{-2.2 cm} \hat {\cal L}   &  \hspace{-1 cm} \equiv & \hspace{-1 cm}  -\rho \left[\frac{\partial{\nu}}{\partial t} + \alpha \frac{\partial{\chi}}{\partial t}
+ \beta \frac{\partial{\eta}}{\partial t}+ \sigma \frac{\partial{s}}{\partial t}+\varepsilon (\rho,s)+
\frac{1}{2} (\vec \nabla \nu + \alpha \vec \nabla \chi +  \beta \vec \nabla \eta +  \sigma \vec \nabla s )^2 \right]
\nonumber \\
\hspace{-2.2 cm} &-&\frac{1}{8 \pi}(\vec \nabla \chi \times \vec \nabla \eta)^2
\nonumber \\
\hspace{-2.2 cm} {\cal L}_{\vec v} &\equiv & \frac{1}{2}\rho (\vec v-\hat{\vec v})^2,
\nonumber \\
\hspace{-2.2 cm} {\cal L}_{\vec B} &\equiv & \frac{1}{8 \pi} (\vec B-\hat{\vec B})^2,
\nonumber \\
\hspace{-2.2 cm} {\cal L}_{boundary} &\equiv & \frac{\partial{(\nu \rho)}}{\partial t} + \vec \nabla \cdot (\nu \rho \vec v ).
\label{Lagactionsimp5}
\enr
The only term containing $\vec v$ is\footnote{${\cal L}_{boundary}$ also depends on
$\vec v$ but being a boundary term is space and time it does not contribute to the derived equations}
 ${\cal L}_{\vec v}$, it can easily be seen that
this term will lead, after we nullify the variational derivative with respect to $\vec v$,
to \ern{vform} but will otherwise
have no contribution to other variational derivatives. Similarly the only term containing $\vec B$
is ${\cal L}_{\vec B}$ and it can easily be seen that
this term will lead, after we nullify the variational derivative, to \ern{Bsakurai2} but will
have no contribution to other variational derivatives. Also notice that the term ${\cal L}_{boundary}$
contains only complete partial derivatives and thus can not contribute to the equations although
it can change the boundary conditions. Hence we see that \ers{lagmul}, \ern{nueq}, \ers{albetaeq} and \er{sigmaeq}
can be derived using the Lagrangian density:
\ber
& & \hat {\cal L}[\alpha,\beta,\chi,\eta,\nu,\rho,\sigma,s] = -\rho [\frac{\partial{\nu}}{\partial t} + \alpha \frac{\partial{\chi}}{\partial t}
+ \beta \frac{\partial{\eta}}{\partial t}+ \sigma \frac{\partial{s}}{\partial t}
\nonumber \\
& & +\  \varepsilon (\rho,s) + \frac{1}{2} (\vec \nabla \nu + \alpha \vec \nabla \chi +  \beta \vec \nabla \eta +  \sigma \vec \nabla s )^2 ]
-\frac{1}{8 \pi}(\vec \nabla \chi \times \vec \nabla \eta)^2
\label{Lagactionsimp6}
\enr
in which $\hat{\vec v}$ replaces $\vec v$ and $\hat{\vec B}$ replaces $\vec B$ in the relevant equations.
Furthermore, after integrating the eight \eqs
(\ref{lagmul},\ref{nueq},\ref{albetaeq},\ref{sigmaeq}) we can insert the potentials $\alpha,\beta,\chi,\eta,\nu,\sigma,s$
into \ers{vform} and (\ref{Bsakurai2}) to obtain the physical quantities $\vec v$ and $\vec B$.
Hence, the general non-barotropic MHD problem is reduced from eight equations
(\ref{Beq},\ref{masscon},\ref{Euler},\ref{Ent}) and the additional constraint (\ref{Bcon})
to a problem of eight first order (in the temporal derivative) unconstrained equations.
Moreover, the entire set of equations can be derived from the Lagrangian density $\hat {\cal L}$.

\section{Further Simplification}

\subsection{Elimination of Variables}

 Let us now look at the three last three equations of (\ref{lagmul}). Those describe three comoving quantities
 which can be written in terms of the generalized Clebsch form given in \ern{vform} as follows:
\ber
& &  \frac{\partial \chi}{\partial t} + (\vec \nabla \nu + \alpha \vec \nabla \chi + \beta \vec \nabla \eta + \sigma \vec \nabla s)
\cdot \vec \nabla \chi = 0
\nonumber \\
& & \frac{\partial \eta}{\partial t} + (\vec \nabla \nu + \alpha \vec \nabla \chi + \beta \vec \nabla \eta + \sigma \vec \nabla s)
\cdot \vec \nabla \eta = 0
\nonumber \\
& & \frac{\partial s}{\partial t} + (\vec \nabla \nu + \alpha \vec \nabla \chi + \beta \vec \nabla \eta + \sigma \vec \nabla s)
\cdot \vec \nabla s = 0
\label{lagmul4}
\enr
Those are algebraic equations for $\alpha, \beta, \sigma$. Which can be solved such that $\alpha, \beta, \sigma$ can be written
as functionals of $\chi,\eta,\nu,s$, resulting eventually in the description of non-barotropic MHD
in terms of five functions: $\nu,\rho,\chi,\eta,s$.
Let us introduce the notation:
\beq
\hspace{-1 cm} \alpha_i \equiv (\alpha, \beta, \sigma), \quad \chi_i\equiv (\chi,\eta,s), \quad
 k_i \equiv -\frac{\partial \chi_i}{\partial t} - \vec \nabla \nu \cdot \vec \nabla \chi_i, \qquad
i\in(1,2,3)
 \label{ali}
\enq
In terms of the above notation \ern{lagmul4} takes the form:
\beq
k_i =\alpha_j \vec \nabla \chi_i \cdot \vec \nabla \chi_j, \qquad j\in(1,2,3)
\label{kieq}
\enq
in which the Einstein summation convention is assumed. Let us define the matrix:
\beq
A_{ij} \equiv  \vec \nabla \chi_i \cdot \vec \nabla \chi_j
\label{Adef}
\enq
obviously this matrix is symmetric since $A_{ij}=A_{ji}$. Hence \er{kieq} takes the form:
\beq
k_i = A_{ij} \alpha_j, \qquad j\in(1,2,3)
\label{kieq2}
\enq
 Provided that the matrix $A_{ij}$ is not singular it has an inverse $A^{-1}_{ij}$ which can be written as:
\beq
\hspace{-1 cm} A^{-1}_{ij}=\left|A\right|^{-1} \left(
\begin{array}{ccc}
 A_{22} A_{33}-A_{23}^2 & A_{13} A_{23}-A_{12} A_{33} & A_{12} A_{23}-A_{13} A_{22} \\
 A_{13} A_{23}-A_{12} A_{33} & A_{11} A_{33}-A_{13}^2 & A_{12} A_{13}-A_{11} A_{23} \\
 A_{12} A_{23}-A_{13} A_{22} & A_{12} A_{13}-A_{11} A_{23} & A_{11} A_{22}-A_{12}^2
\end{array}
\right)
\label{invAdef}
\enq
In which the determinant $\left|A\right|$ is given by the following equation:
\beq
\left|A\right|=
A_{11} A_{22} A_{33}-A_{11} A_{23}^2-A_{22} A_{13}^2 -A_{33} A_{12}^2 +2 A_{12} A_{13} A_{23}
\label{Adet}
\enq
In terms of the above equations the $\alpha_i$'s can be calculated as functionals of $\chi_i,\nu$ as
follows:
\beq
\alpha_i [\chi_i,\nu]= A^{-1}_{ij} k_j.
\label{aleq}
\enq
The velocity \ern{vform} can now be written as:
\beq
\vec v = \vec \nabla \nu + \alpha_i \vec \nabla \chi_i=  \vec \nabla \nu + A^{-1}_{ij} k_j \vec \nabla \chi_i
=\vec \nabla \nu - A^{-1}_{ij}\vec \nabla \chi_i (\frac{\partial \chi_j}{\partial t} + \vec \nabla \nu \cdot \vec \nabla \chi_j).
\label{vform2}
\enq
Provided that the $\chi_i$ is a coordinate basis in three dimensions, we may write:
\beq
\vec \nabla \nu= \vec \nabla \chi_n \frac{\partial \nu}{\partial \chi_n}, \qquad n\in(1,2,3).
\label{nudecom}
\enq
Inserting \ern{nudecom} into \ern{vform2} we obtain:
\ber
\vec v &=& - A^{-1}_{ij}\vec \nabla \chi_i \frac{\partial \chi_j}{\partial t}+
\vec \nabla \nu - A^{-1}_{ij}\vec \nabla \chi_i  \frac{\partial \nu}{\partial \chi_n} \vec \nabla \chi_n \cdot \vec \nabla \chi_j
\nonumber \\
&=& - A^{-1}_{ij}\vec \nabla \chi_i \frac{\partial \chi_j}{\partial t}+
\vec \nabla \nu - A^{-1}_{ij} A_{jn} \vec \nabla \chi_i  \frac{\partial \nu}{\partial \chi_n}
\nonumber \\
&=& - A^{-1}_{ij}\vec \nabla \chi_i \frac{\partial \chi_j}{\partial t}+
\vec \nabla \nu - \delta_{in} \vec \nabla \chi_i  \frac{\partial \nu}{\partial \chi_n}
\nonumber \\
&=& - A^{-1}_{ij}\vec \nabla \chi_i \frac{\partial \chi_j}{\partial t}+
\vec \nabla \nu -  \vec \nabla \chi_n  \frac{\partial \nu}{\partial \chi_n}
\nonumber \\
&=& - A^{-1}_{ij}\vec \nabla \chi_i \frac{\partial \chi_j}{\partial t}
\label{vform3}
\enr
in the above $\delta_{in}$ is a Kronecker delta. Thus the velocity $\vec v [\chi_i]$ is a functional of
$\chi_i$ only and is independent of $\nu$.

\subsection{Lagrangian Density and Variational Analysis}

Let us now rewrite the Lagrangian density $\hat {\cal L}[\chi_i,\nu,\rho]$ given in
\ern{Lagactionsimp6} in terms of the new variables:
\beq
\hspace{-2 cm} \hat {\cal L}[\chi_i,\nu,\rho] = -\rho [\frac{\partial{\nu}}{\partial t} + \alpha_k [\chi_i,\nu] \frac{\partial{\chi_k}}{\partial t}
 +\  \varepsilon (\rho,\chi_3) + \frac{1}{2} \vec v [\chi_i]^2 ]
-\frac{1}{8 \pi}(\vec \nabla \chi_1 \times \vec \nabla \chi_2)^2
\label{Lagactionsimp7}
\enq
Let us calculate the variational derivative of $\hat {\cal L}[\chi_i,\nu,\rho]$ with respect to $\chi_i$ this will result in:
\beq
\hspace{-2 cm} \delta_{\chi_i}\hat {\cal L} = -\rho [  \delta_{\chi_i} \alpha_k  \frac{\partial{\chi_k}}{\partial t} +
\alpha_{\underline{i}}  \frac{\partial \delta \chi_{\underline{i}}}{\partial t}
 +\  \delta_{\chi_i} \varepsilon (\rho,\chi_3) +  \delta_{\chi_i}\vec v \cdot \vec v ]
-\frac{ \vec B} {4 \pi} \cdot  \delta_{\chi_i} (\vec \nabla \chi_1 \times \vec \nabla \chi_2)
\label{delchiLag}
\enq
in which the summation convention is not applied if the index is underlined.
However, due to \ern{vform2} we may write:
\beq
  \delta_{\chi_i}\vec v= \delta_{\chi_i} \alpha_k \vec \nabla \chi_k +   \alpha_{\underline{i}} \vec \nabla \delta \chi_{\underline{i}}.
\label{delchiv}
\enq
Inserting \ern{delchiv} into \ern{delchiLag} and rearranging the terms we obtain:
\ber
 \delta_{\chi_i}\hat {\cal L} &=& -\rho [  \delta_{\chi_i} \alpha_k  (\frac{\partial{\chi_k}}{\partial t}
+ \vec v \cdot \vec \nabla \chi_k )+
\alpha_{\underline{i}}  (\frac{\partial \delta \chi_{\underline{i}}}{\partial t}+ \vec v \cdot \vec \nabla \delta \chi_{\underline{i}})
 +\  \delta_{\chi_i} \varepsilon (\rho,\chi_3) ]
 \nonumber \\
&-& \frac{ \vec B} {4 \pi} \cdot  \delta_{\chi_i} (\vec \nabla \chi_1 \times \vec \nabla \chi_2).
\label{delchiLag2}
\enr
Now by construction $\vec v$ satisfies \ern{lagmul4} and hence $\frac{\partial{\chi_k}}{\partial t}
+ \vec v \cdot \vec \nabla \chi_k  = 0$, this leads to:
\beq
 \delta_{\chi_i}\hat {\cal L} = -\rho \left[  \alpha_{\underline{i}} \frac{d \delta \chi_{\underline{i}}}{d t}
  + \delta_{\chi_i} \varepsilon (\rho,\chi_3) \right] - \frac{ \vec B} {4 \pi} \cdot  \delta_{\chi_i} (\vec \nabla \chi_1 \times \vec \nabla \chi_2).
\label{delchiLag3}
\enq
From now on the derivation proceeds as in \eqs (\ref{delActionchi},\ref{delActioneta},\ref{delActions}) resulting in \eqs
(\ref{albetaeq},\ref{sigmaeq}) and will not be repeated. The difference is that now $\alpha, \beta$ and $\sigma$ are
 not independent quantities, rather they depend through \ern{aleq}
on the derivatives of $\chi_i,\nu$. Thus, \eqs (\ref{delActionchi},\ref{delActioneta},\ref{delActions})
 are not first order equations in time but are second order equations. Now let us calculate the variational derivative
 with respect to $\nu$ this will result in the expression:
\beq
 \delta_{\nu} \hat {\cal L} = -\rho [ \frac{\partial{\delta \nu}}{\partial t} + \delta_{\nu} \alpha_n  \frac{\partial{\chi_n}}{\partial t}]
\label{delnuLag}
\enq
However, $\delta_{\nu} \alpha_k$ can be calculated from \ern{aleq}:
\beq
\delta_{\nu} \alpha_n = A^{-1}_{nj} \delta_{\nu} k_j = - A^{-1}_{nj} \vec \nabla \delta \nu \cdot \vec \nabla \chi_j
\label{delnualeq}
\enq
Inserting the above equation into \ern{delnuLag}:
\beq
 \delta_{\nu} \hat {\cal L} = -\rho [ \frac{\partial{\delta \nu}}{\partial t} - A^{-1}_{nj}  \vec \nabla \chi_j
  \frac{\partial{\chi_n}}{\partial t} \cdot \vec \nabla \delta \nu ] =
  -\rho [ \frac{\partial{\delta \nu}}{\partial t} +  \vec v \cdot \vec \nabla \delta \nu ]=
   -\rho \frac{d{\delta \nu}}{d t}
\label{delnuLag2}
\enq
The above equation can be put to the form:
\beq
 \delta_{\nu} \hat {\cal L} = \delta \nu [\frac{\partial{\rho}}{\partial t} + \vec \nabla \cdot (\rho \vec v )]
-\frac{\partial{(\rho \delta \nu)}}{\partial t}- \vec \nabla \cdot (\rho \vec v \delta \nu )
\label{delnuLag3}
\enq
This obviously leads to the continuity \ern{masscon} and some boundary terms in space and time. The variational
derivative with respect to $\rho$ is trivial and the analysis is identical to the one in \ern{delActionrho} leading
to \ern{nueq}. To conclude this subsection let us summarize the equations of non-barotropic MHD:
\ber
\frac{d \nu}{d t} &=& \frac{1}{2} \vec v^2 - w,
\nonumber \\
\frac{\partial{\rho}}{\partial t} &+& \vec \nabla \cdot (\rho \vec v ) = 0
\nonumber \\
\frac{d \sigma}{dt} &=& T,
\nonumber \\
\frac{d \alpha}{dt} &=& \frac{\vec \nabla \eta \cdot \vec J}{\rho},
\nonumber \\
\frac{d \beta}{dt} &=& -\frac{\vec \nabla \chi \cdot \vec J}{\rho},
\label{equa}
\enr
in which $\alpha,\beta,\sigma,\vec v$ are functionals of $\chi,\eta,s,\nu$ as described above.
It is easy to show as in \ern{dvform2} that those variational equations are equivalent to the physical equations.

It is shown in \cite{Yahnbmhd} tha the Lagrangian density can be written
 standard quadratic form:
\beq
\hspace{-2 cm} \hat {\cal L}[\chi_i,\nu,\rho] = \rho [\frac{1}{2} A^{-1}_{jn} \frac{\partial \chi_j}{\partial t}  \frac{\partial \chi_n}{\partial t}
+\frac{\partial \nu}{\partial \chi_m} \frac{\partial{\chi_m}}{\partial t}-
 \frac{\partial{\nu}}{\partial t} -\  \varepsilon (\rho,\chi_3)]
-\frac{1}{8 \pi}(\vec \nabla \chi_1 \times \vec \nabla \chi_2)^2.
\label{Lagactionsimp8}
\enq
In which $A^{-1}_{jn}$ plays the rule of a "metric". The Lagrangian is thus composed of a kinetic terms which is quadratic in the temporal
derivatives, a "gyroscopic" terms which is linear in the temporal derivative and a potential term which is independent of the temporal derivative.

\section{The Aharonov-Bohm Effect}

Consider an electron moving from A to B (figure \ref{1}) in the
middle we have a magnetic field  $\vec B$ going into the plane
through which the electron is forbidden to pass, hence for the
electron the magnetic field is zero. However, the vector potential
$\vec A$  is not zero, in fact:
\begin{figure}
\vspace{5cm} \includegraphics{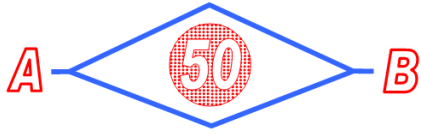} \caption {An electron is moving from point
A to point B. In the middle of the figure we have a confined
magnetic field of 50 Tesla.}
 \label{1}
\end{figure}
\begin{equation}
\vec{B}=\vec{\nabla }\times \vec{A}=0\Rightarrow \vec{A}=\vec{\nabla }\bar{S}
\label{GrindEQ__1_}
\end{equation}
$\vec{\nabla }$ has its standard meaning in vector calculus, $\bar{S}$ is a non single valued function
and its discontinuous variation in value (which we shall call {\it discontinuity} in the following)
$[\bar{S}]$ can be calculated immediately using Stokes theorem:
\begin{equation} \label{GrindEQ__2_}
\Phi =\int \vec{B}\cdot d\vec{S} =\int \vec{\nabla }\times \vec{A}\cdot d\vec{S} =\oint \vec{A}\cdot d\vec{l} =
\oint \vec{\nabla }\bar{S}\cdot d\vec{l} =[\bar{S}]
\end{equation}
Here $\Phi $ is the magnetic flux, the first integral is an area integral and the third is a line integral
in which the trajectory goes around the confined magnetic field. Aharonov and Bohm \cite{AhBo} have shown that $\bar{S}$
is proportional to the phase of the electron wave function. Thus its discontinuity will cause interference at point B.
If the magnetic field is uniform in a cylinder and zero outside the cylinder, the vector potential can be calculated to be:
\begin{equation} \label{GrindEQ__3_}
\vec{A}=A_{\theta } \hat{\theta }=\frac{\Phi }{2\pi \, r} \hat{\theta }\, =
\vec{\nabla }\bar{S}\Rightarrow \bar{S}=\frac{\Phi }{2\pi \, } \theta +\bar{S}_{0} .
\end{equation}
Where $\theta $ is the azimuthal angle.
\begin{figure}
\vspace{6cm} \includegraphics{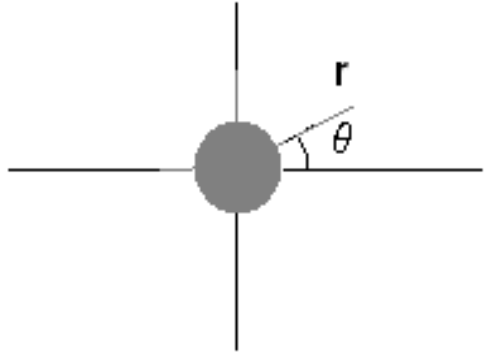} \caption {The azimuth-radius coordinate
system, used to calculate the vector potential. The magnetic field
vanishes except at the gray area.}
 \label{2}
\end{figure}
The main features of the Aharonov - Bohm effect are:
\begin{enumerate}
\item  A domain that is not simply connected due to the presence of a magnetic field, but can be made simply connected by introducing a cut.
Mathematically speaking the domain has a non-trivial fundamental Homotopy group. Two classes of loops exist in the plane; loops that
can be contracted to a point without intersecting the magnetic region and loops that can not.
\item  The electron (or its wave function) do not feel directly the magnetic field -- non locality.
\item  The  potential vector field is a gradient of a non-single valued function.
\item  Gauge freedom is not gone but only limited to single-valued gauges.
\end{enumerate}
The discontinuity $[\bar{S}]$ causes a phase difference of the form $2 \pi \frac{e}{c h}[\bar{S}]$ between the electron's "trajectories".
 It will be shown later that the analogous quantity in MHD do not cause such an effect.
However, it will lead to a new constant of motion of MHD which is the cross helicity per unit of magnetic flux.
It should also be mentioned that according to Bohm's causal interpretation of quantum mechanics there is a quantum - classical correspondence.
According to Bohm \cite{18,19} the phase of a wave function $S$ should be interpreted as a potential of the velocity field $\vec{v}$:
\begin{equation} \label{GrindEQ__4_}
\vec{v}=\frac{1}{m} \vec{\nabla }S
\end{equation}
$m$ is the mass of the particle. However, this correspondence can
go the other way around! If the velocity field has a potential
part it can be interpreted as a phase of a wave function. It will be shown that
this potential function has the topological properties somewhat analogous to a phase even if the wave function does not exist
in the theory under study.

Earlier classical analogues to the Aharonov - Bohm effect were discussed by Berry et al. \cite{Berry}
 which describes a classical analogue to the AB effect in surface waves of swirling water.
  Never the less I would like to highlight the major differences between the approach of Berry et al. and the approach to be described below.
  First in the current paper the classical analogue is related to magnetohydrodynamics while in Berry et al. it is related to
   non magnetic fluids. Second, in the current work one does not assume small perturbations ("waves")
   but rather to results that are valid for any ideal MHD flow. Third one does not assume a "slow change" as in Berry et al.
   paper in which the velocity field is assumed small with respect to the group velocity; in the current work the
    results are correct for any rate of change. Fourth the classical analogue of Berry et al. is not related to any
     vector potential or magnetic flux but they show that a velocity field can play the same rule of a vector
      potential for surface waves interacting with such a velocity field.
 Fifth, the current AB effects are related to topological conservation laws in MHD, while there is no such a relation in \cite{Berry}.

\section{Topological Constants of Motion}

Magnetohydrodynamics is known to have the following two topological constants of motion; one is the magnetic helicity:
\beq
 H_{M} \equiv \int  \vec{B}\cdot \vec{A}d^{3} x,
\label{GrindEQ__21_}
 \enq
 which is known to measure the degree of knottiness of lines of the magnetic field $\vec{B}$ \cite{Woltjer1,Woltjer2}.
 The domain of integration in  \eqref{GrindEQ__21_} is the entire space, obviously regions containing
 a null magnetic field will have a null contribution to the integral. In the above equation $\vec{A}$ is the vector
 potential defined implicitly by \eqref{GrindEQ__1_}.  The other topological constant is the magnetic cross helicity:
\begin{equation} \label{GrindEQ__22b_}
H_{C} \equiv \int  \vec{B}\cdot \vec{v}d^{3} x,
\end{equation}
characterizing the degree of cross knottiness of the magnetic field and vortex lines.
The domain of integration in  \eqref{GrindEQ__22_} is the magnetohydrodynamic flow domain. As noticed before this
second topological constant of motion is only constant for barotropic or incompressible MHD.
Notice that in non-barotropic MHD:
\begin{equation} \label{GrindEQ__22c_}
\frac{dH_{C}}{dt}= \int T  \vec{\nabla} S \cdot \vec{B} d^{3} x,
\end{equation}
hence generally speaking cross helicity is not conserved.

\section {Non-Barotropic Cross Helicity}

 A clue on how to define cross helicity
 for non-barotropic MHD can be obtained from
the variational analysis described in the previous sections.

Let us now write the cross helicity given in \ern{GrindEQ__22_} in terms of \ern{Bsakurai2} and \ern{vform},
this will take the form:
\begin{equation} \label{GrindEQ__22d_}
H_{C} = \int d \Phi [\nu] + \int d \Phi \oint \sigma d s
\end{equation}
in which: $d\Phi =\vec{B}\cdot d\vec{S}=\vec{\nabla }\chi \times \vec{\nabla }\eta \cdot d\vec{S}=d\chi \, d\eta $ and the closed line
integral is taken along a magnetic field line. $d\Phi$ is a magnetic flux element which is
comoving according to \ern{Beq} and $d\vec{S}$ is an infinitesimal area element. Although the
cross helicity is not conserved for non-barotropic flows, looking at the right hand side we see
that it is made of a sum of two terms. One which is conserved as both $d\Phi$ and $[\nu]$ are comoving (see \ern{mplicated nueqc})
and one which is not. This suggests the following definition for the non barotropic cross helicity $H_{CNB}$:
\begin{equation} \label{GrindEQ__22e_}
H_{CNB} \equiv \int d \Phi [\nu] = H_{C} -  \int d \Phi \oint \sigma d s
\end{equation}
Which can be written in a more conventional form:
\begin{equation} \label{GrindEQ__22e1_}
H_{CNB} = \int  \vec{B} \cdot \vec v_t  d^{3} x
\end{equation}
where the topological velocity field is defined as follows:
\begin{equation} \label{vt_}
 \vec v_t \equiv \vec v - \sigma \vec \nabla s
\end{equation}
It should be noticed that $H_{CNB}$ is conserved even for an MHD not satisfying
the Sakurai topological constraint given in \ern{Bsakurai2}, provided that we have
a field $\sigma$ satisfying the equation $\frac{d \sigma}{dt} = T$. Thus the non barotropic cross helicity
conservation law:
\begin{equation} \label{HCNBcon}
\frac{d H_{CNB}}{dt} = 0,
\end{equation}
is more general than the variational principle described by \ern{Lagactionsimp8} as follows from a
direct computation using  \eqs (\ref{Beq},\ref{masscon},\ref{Euler},\ref{Ent}). Also notice that for a constant
specific entropy $S$ we obtain $H_{CNB}=H_{C}$ and the non-barotropic cross helicity reduces to the standard barotropic cross helicity.
To conclude we introduce also a local topological conservation law in the spirit of \cite{Yah2} which is the
non barotropic cross helicity per unit of magnetic flux. This quantity which is equal to the discontinuity of $\nu$  is conserved and
 can be written as a sum of the barotropic cross helicity per unit flux and the closed line integral of $s d \sigma$ along a magnetic field line:
\begin{equation} \label{loc_}
[\nu]= \frac{dH_{CNB}}{d \Phi} = \frac{dH_{C}}{d \Phi} + \oint  s d \sigma.
\end{equation}

\section {Local Cross Helicities}

\begin{figure}
\vspace{5cm}
\includegraphics{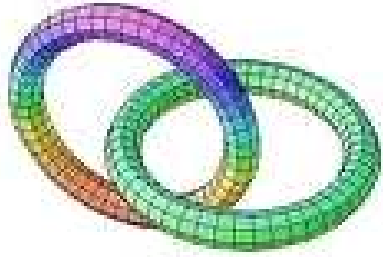}
\caption{Knotted magnetic field lines with none zero magnetic helicity and a non-single valued $\zeta $.}
 \label{3}
\end{figure}

 Let us write the topological constants given in  \eqref{GrindEQ__21_} and  \eqref{GrindEQ__22_}
 in terms of the magnetohydrodynamic potentials $\chi,\eta,\nu,\rho$ introduced in previous sections.
 First let us combine   \eqref{GrindEQ__1_} with  \eqref{Bsakurai2} to obtain the equation:
\begin{equation} \label{GrindEQ__23_}
\vec{\nabla }\times (\vec{A}-\chi \vec{\nabla }\eta )=0,
\end{equation}
 this leads immediately to the result:
\begin{equation} \label{GrindEQ__24_}
\vec{A}=\chi \vec{\nabla }\eta +\vec{\nabla }\zeta ,
\end{equation}
 in which $\zeta$ is some function. Let us now calculate the scalar product $\vec{B}\cdot \vec{A}$:
\begin{equation} \label{GrindEQ__25_}
\vec{B}\cdot \vec{A}=(\vec{\nabla }\chi \times \vec{\nabla }\eta )\cdot \vec{\nabla }\zeta .
\end{equation}
We can define a local vector basis: $(\vec \nabla \chi ,\vec \nabla \eta ,\vec \nabla \mu)$
based on the magnetic field lines. Here, in addition to $\chi,\eta$, we have added another coordinate the magnetic metage
$\mu$ which parameterizes the distance along the magnetic field lines \cite{YaLy}. $\vec{\nabla }\zeta$ can thus be written as:
\begin{equation} \label{GrindEQ__26_}
\vec{\nabla }\zeta =\frac{\partial \zeta }{\partial \chi } \vec{\nabla }\chi +
\frac{\partial \zeta }{\partial \mu } \vec{\nabla }\mu +\frac{\partial \zeta }{\partial \eta } \vec{\nabla }\eta.
\end{equation}
 Hence we can write:
\begin{equation} \label{GrindEQ__27_}
\vec{B}\cdot \vec{A}=\frac{\partial \zeta }{\partial \mu } (\vec{\nabla }\chi \times \vec{\nabla }\eta )\cdot \vec{\nabla }\mu =
\frac{\partial \zeta }{\partial \mu } \left|\frac{\partial (\chi ,\eta ,\mu )}{\partial (x,y,z)}\right| .
\end{equation}
 Let us think of the entire space outside the magnetohydrodynamic domain as containing low density matter. In this case we can
 define the metage $\mu $ over the entire portion of space containing magnetic field lines and the integration domain of
\eqref{GrindEQ__21_} and  \eqref{GrindEQ__22_} coincide. Now we
can insert  \eqref{GrindEQ__27_}  into \eqref{GrindEQ__21_} to obtain the expression:
\begin{equation} \label{GrindEQ__28_}
H_{M} =\int  \frac{\partial \zeta }{\partial \mu } d\mu d\chi d\eta .
\end{equation}
  We can think about the magnetohydrodynamic domain as composed of thin closed tubes of magnetic lines each labelled by $(\chi ,\eta )$.
 Performing the integration along such a thin tube in the metage direction results in:
\begin{equation} \label{GrindEQ__29_}
\oint _{\chi ,\eta } \frac{\partial \zeta }{\partial \mu } d\mu =[\zeta ]_{\chi ,\eta } ,
\end{equation}
 in which $[\zeta ]_{\chi ,\eta } $ is the discontinuity of the function $\zeta $ along its cut, i.e., the shift in value going around the path.
 Thus a thin tube of magnetic lines in which $\zeta $ is single valued does not contribute to the magnetic helicity integral.
 Inserting  \eqref{GrindEQ__29_} into  \eqref{GrindEQ__28_} will result in:
\begin{equation} \label{GrindEQ__30_}
H_{M} =\int  [\zeta ]_{\chi ,\eta } d\chi d\eta =\int  [\zeta ]d\Phi ,
\end{equation}
in which we have used $d\Phi =\vec{B}\cdot d\vec{S}=\vec{\nabla }\chi \times \vec{\nabla }\eta \cdot d\vec{S}=d\chi \, d\eta $. Hence:
\begin{equation} \label{GrindEQ__31_}
[\zeta ]=\frac{dH_{M} }{d\Phi } ,
\end{equation}
the discontinuity of $\zeta$ is thus the density of magnetic helicity per unit of magnetic flux in a tube.
We deduce that the Sakurai representation does not entail zero magnetic helicity, rather it is perfectly consistent
 with non zero magnetic helicity as was demonstrated above. Notice however, that the topological structure of the magnetohydrodynamic
 flow constrains the gauge freedom which is usually attributed to a  vector potential $\vec{A}$ and limits it to single valued functions.
 Moreover, while the choice of $\vec{A}$ is arbitrary since one can add to $\vec{A}$ an arbitrary gradient of a single valued function
  which may lead to different choice of $\zeta $, the discontinuity value $[\zeta ]$ is not arbitrary and has meaning as given above.
  The main features of this novel "Magnetic Aharonov-Bohm effect" are similar to the features of the standard Aharonov-Bohm effect.
\begin{enumerate}
\item  A domain that is not simply connected, since the internal magnetic flux is
 knotted inside the external magnetic flux line (see figure \ref{3}).
\item  The external magnetic field line does not touch the internal flux yet the
$\zeta $ function is not single valued due to that line - non locality.
\item  The  potential vector field has a gradient of a non-single valued function part.
\item  Gauge freedom is not gone but only limited to single-valued gauges.
\end{enumerate}
One should notice that $[\zeta ]_{\chi ,\eta }$ is a conserved quantity, this can easily seen by integrating $\vec \nabla \zeta$
along a closed path at the intersection of the $\chi$ and $\eta$ surfaces, this path is in fact a magnetic field line:
\beq
[\zeta ]_{\chi ,\eta }= \oint_{\chi ,\eta } \vec \nabla \zeta \cdot d \vec r =
 \oint_{\chi ,\eta } \left( \vec A - \chi \vec \nabla \eta  \right) \cdot d \vec r =
  \oint_{\chi ,\eta } \vec A \cdot d \vec r =  \int \vec B \cdot d \vec S
\label{zetajcon}
 \enq
in which we have used Stokes theorem for the last equality sign. Obviously the flux cannot escape or enter into
this field line since all magnetic field lines are comoving, thus not only the magnetic helicity is conserved but
also $[\zeta ]_{\chi ,\eta }$ which is the magnetic helicity per unit flux.

The reader now is reminded of \eqref{loc_}:
\begin{equation} \label{GrindEQ__36_}
[\nu ]=\frac{dH_{CNB}}{d\Phi} ,
\end{equation}
 the discontinuity of $\nu $ is thus the density of cross helicity per unit of magnetic flux.
 We deduce that a flow with null non-barotropic cross helicity will have a single valued $\nu $ function alternatively,
 a non single valued $\nu $ may entail a non zero non-barotropic cross helicity. Furthermore, from  \eqref{mplicated nueqc} it is obvious that:
\begin{equation} \label{GrindEQ__37_}
\frac{d[\nu ]}{dt} =0.
\end{equation}
 We conclude that not only is the non-barotropic magnetic cross helicity conserved as an integral quantity of the entire magnetohydrodynamic
 domain but also the (local) density of cross helicity per unit of magnetic flux is a conserved quantity as well.
 
The main features of this novel "Cross Aharonov-Bohm effect" are similar to the features of the standard Aharonov-Bohm effect:
\begin{enumerate}
\item  A domain that is not simply connected, since the internal magnetic flux is knotted inside the external topological stream line.
\item  The topological stream line does not touch the internal flux yet the $\nu$ function is not single valued due to that line - non locality.
\item  The  topological velocity field has a gradient of a non-single valued function part, this part is interpreted as a phase according to
Bohm's causal interpretation correspondence see  \eqref{GrindEQ__4_}.
\end{enumerate}

\section{Possible Application}

 In his important review paper "Physics of magnetically confined plasmas" A. H. Boozer \cite{Boozer} states that: "A spiky
current profile causes a rapid dissipation of energy relative to magnetic helicity. If the evolution of a
magnetic field is rapid, then it must be at constant helicity." This will also be true also for the magnetic helicity per unit flux.
The application of the "Magnetic Aharonov-Bohm effect" is expected to be important in understanding the dynamics of magnetically
confined plasmas and the problem of controlled fusion. Usually topological conservation laws are used in order to deduce
lower bounds on the "energy" of the flow.
 Those bounds are only approximate in non ideal flows but due to their topological nature simulations show that
 they are approximately conserved even when the "energy" is not. For example it is easy to show that the "energy"
 is bounded from below by the magnetic helicity as follows:
 \begin{equation}
H_{M} = \int  \vec{B}\cdot \vec{A}d^{3} x \leq \frac{1}{2}\int  \left( \vec{B}^2 + \vec{A}^2 \right) d^{3} x,
 \label{Globhelboundm}
\end{equation}
We point out that the Cauchy-Schwarz inequality also holds:
\begin{equation}
H_{M} = \int  \vec{B}\cdot \vec{A}d^{3} x \leq \sqrt{\int  \vec{A}^2 d^{3} x}\sqrt{\int \vec{B}^2 d^{3} x},
 \label{SchwarzHm}
\end{equation}

In this sense a configuration with a highly complicated topology is more stable since its "energy" is bounded from below.
However, the above constraint is only global. Using the magnetic AB effect one may deduce a more local constraint.
Consider a magnetic flux tube of a cross section $\Delta S$  in which the magnetic field is almost constant in this tube:
\beq
H_{M} =\int d^{3} x \vec B \cdot \vec A \simeq [\zeta ]B\Delta S
\label{Fluxtubem}
\end{equation}
Hence in this flux tube we deduce the lower bounds:
\beq
[\zeta ]B\Delta S\le \frac{1}{2} \int d^3 x \left(\vec A ^2 +\vec B^2 \right)
\label{GlobhelboundFluxtubem}
\end{equation}
\beq
[\zeta ]B\Delta S\le \sqrt{\int  \vec{A}^2 d^{3} x}\sqrt{\int \vec{B}^2 d^{3} x}
\label{GlobhelboundFluxtubem2}
\end{equation}
Writing the above equation using: ${\rm d}^{{\rm 3}} {\rm x}=\Delta S dl$, in which $dl$
is a line element along the flux tube we obtain the local bounds:
\beq
[\zeta ] \le \frac{1}{2B} \int dl\left(\vec A ^2 +\vec B^2 \right)
\label{LochelboundFluxtubem}
\end{equation}
\beq
[\zeta ] \le \frac{1}{B} \sqrt{\int  \vec{A}^2  dl}\sqrt{\int \vec{B}^2  dl} =  \sqrt{L \int  \vec{A}^2  dl} 
\label{LochelboundFluxtubem2}
\end{equation}
in which $L$ is the length of the flux tube. This is a much more stringent bound than the global bound of magnetic helicity.
It may be suggested based on the analysis presented to create in Tokamak devices flows with high amount of local magnetic helicity per unit flux
(which is the same as the discontinuity of the Aharonov-Bohm phases),
those flows are expected to be more stable than flows in which there is no sufficient local magnetic helicity per unit flux.

 A similar analysis can be done for non-barotropic cross helicity per unit flux. It is easy to show that the "energy"
 is bounded from below by the cross helicity as follows:
 \begin{equation}
H_{CNB} = \int  \vec{B}\cdot \vec{v_t}d^{3} x \leq \frac{1}{2}\int  \left( \vec{B}^2 + \vec{v_t}^2 \right) d^{3} x,
 \label{Globhelbound}
\end{equation}
\begin{equation}
H_{CNB} = \int  \vec{B}\cdot \vec{v_t}d^{3} x \leq \sqrt{\int  \vec{v_t}^2 d^{3} x}\sqrt{\int \vec{B}^2 d^{3} x},
 \label{Globhelbound2}
\end{equation}
In this sense a configuration with a highly complicated topology is more stable since its energy is bounded from below.
However, the above constraint is only global. Using the cross AB effect one may deduce a more local constraint.
Consider a magnetic flux tube of a cross section $\Delta S$  in which the magnetic field is almost constant in this tube:
\beq
H_{CNB} =\int d^{3} x \vec B \cdot \vec v_t \simeq [\nu ]B\Delta S
\label{Fluxtube}
\end{equation}
Hence in this flux tube we deduce the lower bounds:
\beq
[\nu ]B\Delta S\le \frac{1}{2} \int d^3 x \left(\vec v_t^2 +\vec B^2 \right)
\label{GlobhelboundFluxtube}
\end{equation}
\beq
[\nu ]B\Delta S\le \sqrt{\int  \vec{v_t}^2 d^{3} x}\sqrt{\int \vec{B}^2 d^{3} x},
\label{GlobhelboundFluxtube2}
\end{equation}
Writing the above equation using: ${\rm d}^{{\rm 3}} {\rm x}=\Delta S dl$, in which $dl$
is a line element along the flux tube we obtain the local bounds:
\beq
[\nu ] \le \frac{1}{2B} \int dl\left(\vec v_t ^2 +\vec B^2 \right)
\label{LochelboundFluxtube}
\end{equation}
\beq
[\nu ] \le \frac{1}{B}\sqrt{\int  \vec{v_t}^2 dl}\sqrt{\int \vec{B}^2 dl} = \sqrt{L \int  \vec{v_t}^2 dl}
\label{LochelboundFluxtube2}
\end{equation}
This is a much more stringent bound than the global bound of cross Helicity.
Notice, however, that there is a difference in the consequence of magnetic and cross-helicity conservation in non ideal magnetohydrodynamics.
The rapid dissipation of the energy relative to magnetic helicity is possible due to the difference in the turbulent cascade
 of these values in the flows. The situation in the case of non-barotropic cross helicity is unknown at present as its a new constant of motion.
 Both values are conserved in ideal flows.  However, nothing prevents the cascade of the energy to the
 small scales where it can dissipate by means of the molecular diffusivity.
 The cascade of the magnetic helicity is subjected to realizability condition: $|H_m(k)|k \le 2 E_m(k)$
 (see \cite{Moffatt78} chapter 11 equation 11.37). This complicates the cascade of the magnetic helicity to the small scales.
In other words, in the small-scales the components of the field contribute to the helicity as $\frac{\vec B^2}{k}$ and to the energy as $\vec B^2$.
Therefore, the changes in the small-scales harmonics of the field make a much smaller contribution to the helicity changes
 than to the energy of the field. The same conclusion is reached by Arnold \& Khesin \cite{ArKh} who claim that the
 dissipation of magnetic helicity is proportional to the resistivity square making the above constraint valid for
 the case of small resistivity. Notice, however, that Taylor \cite{Taylor} points out that in extremely violent MHD flows
local topological constraints do not hold, and therefore even $[\zeta ]$ is not conserved but only global magnetic helicity is conserved
(see also Biskamp \cite{Biskamp}). A generalization of the force-free Taylor's relaxation states studied in laboratory experiments
 (in spheromaks) that become non force-free in the self-gravitating stellar case were obtained by Duez and Mathis \cite{DuMa}.
However, Braithwaite \cite{Braithwaite} studying non-axisymmetric magnetic equilibria in stars has presented numerical simulations of
the formation of stable equilibria from turbulent initial conditions and demonstrated the existence
of non-axisymmetric equilibria consisting of twisted flux tubes lying horizontally below
the surface of the star, meandering around the star in random patterns, he concluded that in
configurations with more than one flux tube, each tube may have
either positive or negative local magnetic helicity; although whether negative or positive has clear implications
for the global helicity it has no effect on the stability. And also stable zero-global helicity equilibria
are possible (but with non zero local $[\zeta ]$). The magnetic helicity conservation was found to play a major
rule for the asymmetry of sunspot cycles due to the effect
of magnetic helicity on the nonlinear surface-shear shaped dynamo \cite{Pipin}.
As for cross helicity, even global cross helicity is not conserved in turbulent fluids such as the solar convection zone
and its balance is controlled by local processes \cite{SolPhys}. This being said it remains to be seen what are the consequences 
of non-barotropic cross helicity both local and global.

It may be suggested based on the analysis presented to create in Tokamak devices flows with high amount of local magnetic helicity per unit flux
and non-barotropic cross helicity per unit flux (which is the same as the discontinuity of the Aharonov-Bohm phases),
those flows are expected to be more stable than flows in which there is no sufficient local magnetic helicity per unit flux.

\section{Example of a Magnetic Aharonov-Bohm \-Phase}

\begin{figure}
\vspace{8cm}
\includegraphics{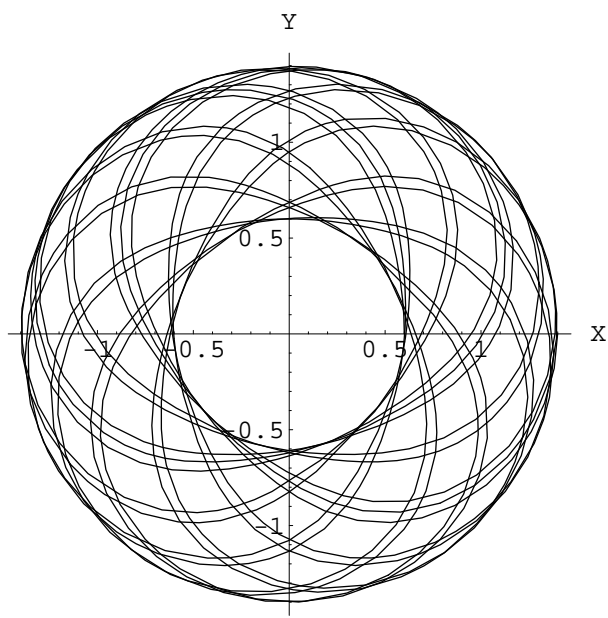}
\caption {A numerically integrated field line assuming that $\Psi = r + r^3,\alpha=1$ and starting from the point
$R=0.6, \phi=0, z=0$. The plot shows twenty rotations.}
\label{fieldline}
\end{figure}

In this penultimate section I would like to give a concrete example of the
calculation of the magnetic Aharonov-Bohm phase \cite{YaLy}. Consider a magnetohydrodynamic flow of uniform density $\rho$.
Furthermore assume (following Moffatt \cite{Moffatt0}) that
the flow contains a vector potential:
\beq
\vec A = \vec \nabla \Psi \times \vec \nabla \phi + \alpha \Psi \vec \nabla \phi =
   \frac{1}{R} \partial_R \Psi \hat{z} - \frac{1}{R} \partial_z \Psi \hat{R} + \frac{\alpha \Psi}{R}\hat{\phi},
   \qquad \vec \nabla \phi = \frac{\hat{\phi}}{R},
\label{ASK}
\enq
in which as in the previous section $R,\phi,z$ are the standard cylindrical coordinates,
$\hat{R},\hat{\phi},\hat{z}$ are the corresponding unit vectors, $\alpha$ is constant and $\Psi = \Psi(R,z)$
 is an arbitrary function
of $R$ and $z$. The magnetic field can be calculated using \ern{GrindEQ__1_} to be:
\beq
\vec B = \frac{\alpha}{R} \partial_R \Psi \hat{z} - \frac{\alpha}{R} \partial_z \Psi \hat{R} -
\frac{D^2 \Psi}{R}\hat{\phi}.
\label{BSK}
\enq
In which according to Moffatt \cite{Moffatt0} the operator $D^2$ is defined as:
\beq
D^2=\partial_z^2+R \partial_R (\frac{1}{R} \partial_R).
\label{Dsq}
\enq
Obviously both $\vec A$ and $\vec B$ lie on $\Psi$ surfaces since:
\beq
\vec \nabla \Psi \cdot \vec A = \vec \nabla \Psi \cdot \vec B =0.
\label{ABsurfaces}
\enq
Let us define the variable $r$:
\beq
r = \sqrt{z^2+(R-1)^2}.
\enq
And let us assume that $\Psi = \Psi (r)$ is a function of $r$. In this case surfaces of constant $\Psi$
are nested tori. The magnetic field is assumed to be confined between the tori $0 \leq r \leq a$ in which $a$
is an arbitrary number such that $0 < a < 1$. A depiction of a $R-z$ cross section of the  nested tori is given in
figure \ref{nesttori} below.
\begin{figure}
\vspace{12cm}
\includegraphics{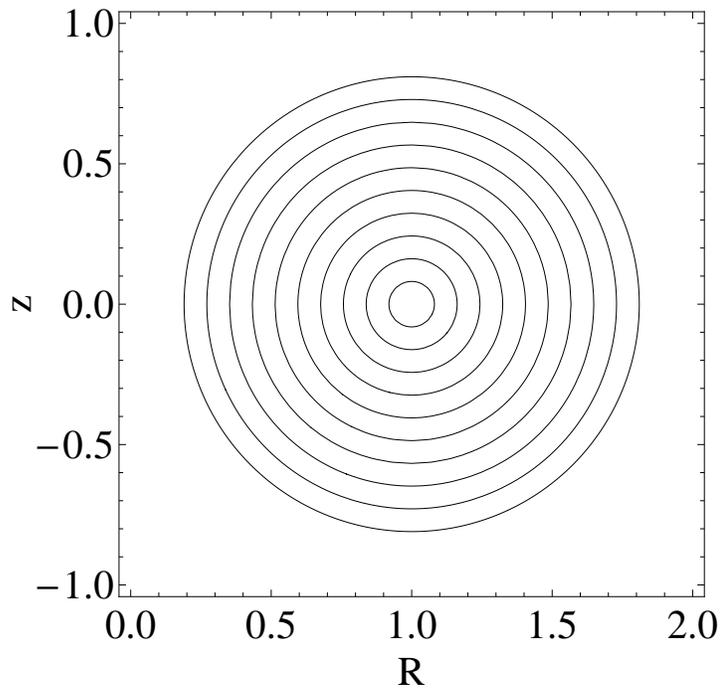}
\caption {$R-z$ cross section of the nested tori.}
\label{nesttori}
\end{figure}
A typical field line of the magnetic field given in \ern{BSK} is  self knotted in the sense of Moffatt \cite{Moffatt0}
as is evident from figure \ref{fieldline}.

\begin{figure}
\vspace{6cm}
\includegraphics{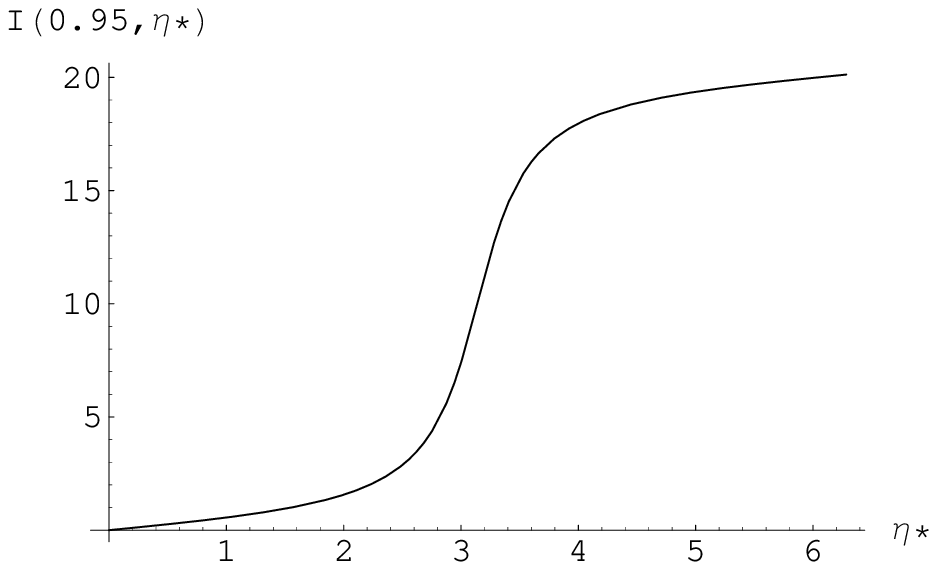}
\caption {$I(r,\eta^{*})$ for $r=0.95$.}
\label{Iplot}
\end{figure}

\begin{figure}
\vspace{6cm}
\includegraphics{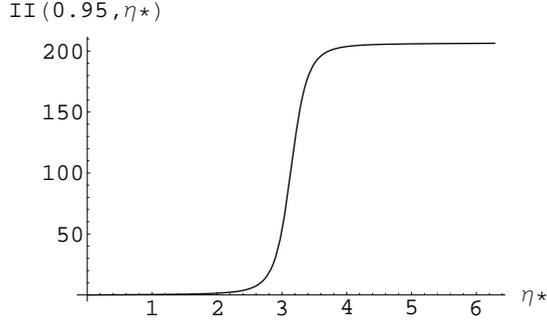}
\caption {$II(r,\eta^{*})$ for $r=0.95$.}
\label{IIplot}
\end{figure}

Next, we define two functions with simple cuts $\phi^{*}$ and $\eta^{*}$.
In which $\eta^{*}$ can be considered as an angle varying over the small circle of the torus,
while $\phi^{*}$ can be considered as an angle varying over the large circle of the torus.
Hence $\phi^{*}=\phi$ is just the standard azimuthal angle and $\eta^{*}$ can be defined as:
\beq
\eta^{*} = \arctan{\frac{z}{R-1}}.
\label{etast}
\enq
Obviously the $\Psi$ surfaces are also the $\lambda$ surfaces. Therefore we can calculate $\chi$
using equation 4.31 of \cite{YaLy} were we calculate the magnetic flux into the surface between the degenerate torus
$r=0$ and any other torus given by some value of $\Psi$. There are two ways to do this but it seems
that the simpler way is to take the surface which is perpendicular to $\hat{\eta^{*}}$ which
is a unit vector in the $\eta^{*}$ direction. Hence we obtain:
\beq
\chi=\frac{1}{2 \pi} \int \vec B \cdot d \vec S = \frac{1}{2 \pi} \oint \vec A \cdot d \vec l
= \frac{1}{2 \pi} \int_0^{2 \pi}  A_\phi R d \phi = A_\phi R = \alpha \Psi,
\label{chical}
\enq
in the above we assumed that $\Psi(0)=0$. Let us now calculate the function $\eta$ by solving \ern{Bsakurai2}.
It is easy to show that $\eta$ is of the form:
\beq
\eta = \phi + C(z,R),
\label{eta1}
\enq
in which $C(z,R)$ is a solution of:
\beq
B_{\phi} = \partial_z \chi \partial_R C -\partial_z \chi \partial_z C.
\enq
Writing the above equation in terms of $r,\eta^{*}$ coordinates we obtain:
\ber
  & & -  \frac{1}{1+r \cos{\eta^{*}}}  \left( \Psi'' + \frac{1}{1+r \cos{\eta^{*}}} \frac{ \Psi'}{r}\right)
   =  -\frac{\alpha \Psi'}{r} \partial_{\eta^{*}} C,
 \nonumber \\
 & & \Psi' \equiv \frac{d\Psi}{dr}, \quad \Psi'' \equiv \frac{d^2\Psi}{dr^2}.
\enr
$C$ can be integrated to yield the solution:
\beq
C = \frac{1}{\alpha}\left[ \frac{r \Psi''}{\Psi'} I(r,\eta^{*}) + II(r,\eta^{*})\right],
\label{Csolution}
\enq
in which:
\ber
& & I(r,\eta^{*}) \equiv \int \frac{d \eta^{*}}{1+r \cos{\eta^{*}}}
\nonumber \\
&=& \frac{2}{\sqrt{1-r^2}} \left[\arctan(\sqrt{\frac{1-r}{1+r}} \tan(\frac{\eta^{*}}{2}))
+
\left\{%
\begin{array}{ll}
   0, & 0 \leq \eta^{*} < \pi \\
   \pi, & \pi \leq \eta^{*} < 2 \pi. \\
\end{array}
\right.\right]
\enr
and
\beq
II(r,\eta^{*}) \equiv \int \frac{d \eta^{*}}{(1+r \cos{\eta^{*}})^2}
= \frac{I(r,\eta^{*})}{1-r^2}- \frac{r \sin \eta^{*} }{(1-r^2)(1+r \cos{\eta^{*}})}.
\enq
Plots of $I(r,\eta^{*})$ and $II(r,\eta^{*})$ are given in figures \ref{Iplot} and \ref{IIplot} respectively.
Obviously $I(r,\eta^{*})$ and $II(r,\eta^{*})$ are non-single valued functions. Their discontinuity values
across the cut are given by:
\beq
[I(r,\eta^{*})]=\frac{2 \pi}{\sqrt{1-r^2}}, \qquad [I(r,\eta^{*})]=\frac{2 \pi}{(1-r^2)^{3/2}} \ .
\enq
Therefore $C(r,\eta^{*})$ is also a non-single valued function. Using \ern{Csolution} we obtain
the following discontinuity value of $C(r,\eta^{*})$ across the cut:
\beq
[C] = \frac{2 \pi}{\alpha \sqrt{1-r^2}}\left( \frac{r \Psi''}{\Psi'} + \frac{1}{1-r^2}\right).
\label{Ccut}
\enq
It remains to calculate the magnetic Aharonov-Bohm function $\zeta$, this can be done using \ern{GrindEQ__24_}.
Inserting into \ern{GrindEQ__24_} the value of $\eta$ given in \ern{eta1}, we obtain:
\beq
\vec A =  \chi \vec \nabla \eta + \vec \nabla \zeta = \frac{\alpha \Psi}{R} \hat{\phi}
+ \alpha \Psi \vec \nabla C + \vec \nabla \zeta.
\label{vecpotIL}
\enq
Taking into account \ern{ASK} in the above equation leads to:
\beq
\vec \nabla \zeta =\frac{1}{R} \partial_R \Psi \hat{z} - \frac{1}{R} \partial_z \Psi \hat{R}-
 \alpha \Psi \vec \nabla C.
\label{zeteq}
\enq
The above equation implies that the magnetic Aharonov-Bohm function $\zeta$ is a function of $R,z$ (or $r,\eta^*$) only.
Writing \ern{zeteq} in terms of the $r,\eta^*$ coordinates we arrive at a set of two equations:
\beq
\frac{1}{r} \partial_{\eta^*} \zeta = - \frac{\alpha \Psi}{r} \partial_{\eta^*} C
+ \frac{\Psi'}{1+r \cos{\eta^{*}}} ,
\qquad \partial_r \zeta = - \alpha \Psi \partial_r C.
\label{zeteq2}
\enq
Solving \ers{zeteq2} we arrive at the solution:
\beq
\zeta (r,\eta^{*})= r \Psi' I(r,\eta^{*}) - \alpha \Psi C
= r I(r,\eta^{*}) \left( \Psi' - \frac{\Psi \Psi''}{\Psi'} \right) - \Psi II(r,\eta^{*})
\label{zetsol}
\enq
Obviously the magnetic Aharonov-Bohm function $\zeta(r,\eta^{*})$ is  a non-single valued function with
the following discontinuity value across the cut:
\beq
[\zeta (r,\eta^{*})] = \frac{2 \pi}{\sqrt{1-r^2}}
\left( r ( \Psi' - \frac{\Psi \Psi''}{\Psi'} ) - \frac {\Psi}{1-r^2} \right)
\label{zetcut}
\enq
Let us calculate the magnetic helicity of the field using \ern{GrindEQ__30_}, \ern{zetcut} and \ern{chical},
we arrive at the result:
\ber
{\cal H}_M &=& \int [\zeta] d\Phi =  \int_0^a [\zeta] 2 \pi \alpha \Psi' dr
\nonumber \\
&=& (2 \pi)^2 \alpha \int_0^a
\frac{dr}{\sqrt{1-r^2}} \left( r ( (\Psi')^2 - \Psi \Psi'' ) - \frac {\Psi \Psi'}{1-r^2} \right).
\label{maghel3IL}
\enr
A direct calculation using \ern{GrindEQ__21_} will yield an identical result. This integral
can be calculated either analytically or numerically for any reasonable function $\Psi(r)$.
For example taking $\Psi(r) = r + r^3$ and  $a=0.9$ we calculated ${\cal H}_M$ numerically
and obtained ${\cal H}_M= -4.6167 (2 \pi)^2 \alpha $. Thus the magnetic Aharonov-Bohm phase
is calculated in a specific example and the magnetic helicity is derived from that phase.

\section{Conclusion}

 To conclude, it is shown that there are two inherent Aharonov - Bohm effects in magnetohydrodynamics.
 In each case a magnetic flux induces a "phase" on quantities that do not come under the influence of the magnetic field directly.
 Those quantities include the topological velocity fields and magnetic fields. Jumps in the phases $\zeta ,\nu $ circumnavigating a closed contour quantify the presence of a topological defect in the vector potential field and the topological velocity field, respectively, and
  these are associated with the two conserved quantities in non-barotropic magnetohydrodynamics, the magnetic helicity and the non-barotropic cross helicity.
   The quantity $\nu$ is useful for introducing a very efficient variational
 principle for MHD which is given in terms of only five independent functions for non-stationary flows. Moreover, the discontinuities $[\nu ]$
 and $[\zeta ]$  which are the non-barotropic cross helicity per unit of magnetic flux and the magnetic helicity per unit of magnetic flux respectively, are conserved
 quantities along the non-barotropic MHD flow. The application of the
  "Magnetic Aharonov-Bohm effect" and the  "Non-Barotropic Cross Aharonov-Bohm effect" may be important in understanding the dynamics of magnetically confined plasmas and the problem of controlled fusion.


\section*{References}

\begin {thebibliography}9
\bibitem{Woltjer1}
Woltjer L, . 1958a Proc. Nat. Acad. Sci. U.S.A. 44, 489-491.
\bibitem{Woltjer2}
Woltjer L, . 1958b Proc. Nat. Acad. Sci. U.S.A. 44, 833-841.
\bibitem {Mobbs}
Mobbs, S.D. (1981) 'Some vorticity theorems and conservation laws for non-barotropic fluids', Journal of Fluid Mechanics, 108, pp. 475–483. doi: 10.1017/S002211208100222X.
\bibitem {Moffatt0}
Moffatt H. K. J. Fluid Mech. 35 117 (1969)
\bibitem{Yahalomhel}
A. Yahalom, "Helicity Conservation via the Noether Theorem" J. Math. Phys. 36, 1324-1327 (1995).
[Los-Alamos Archives solv-int/9407001]
\bibitem{Padhye1}
N. Padhye and P. J. Morrison, Phys. Lett. A 219, 287 (1996).
\bibitem{Padhye2}
N. Padhye and P. J. Morrison, Plasma Phys. Rep. 22, 869 (1996).
\bibitem {YaLy}
Yahalom A. and Lynden-Bell D., "Simplified Variational Principl\-es for Baro\-tropic Magneto\-hydro\-dynamics," (Los-Alamos Archives-
physics/0603128) {\it Journal of Fluid Mechanics}, Vol.~607, 235--265, 2008.
\bibitem {Webb1}
Webb et al. 2014a, J. Phys. A, Math. and theoret., Vol. 47, (2014), 095501 (33pp)
\bibitem {Webb2}
Webb et al. 2014b: J. Phys A, Math. and theoret. Vol. 47, (2014), 095502 (31 pp).
\bibitem {Webb3}
Webb, G.M., and Anco, S.C. 2016, Vorticity and Symplecticity in Multi-symplectic Lagrangian gas dynamics, J. Phys A, Math. and Theoret., 49, No. 7, Feb. 19 issue, (2016) 075501 (44pp), doi:10.1088/1751-8113/49/7/075501.
\bibitem {Webb4}
Webb, G. M., McKenzie, J.F. and Zank, G. P. 2016, J. Plasma Phys., 81, doi:10.1017\-/S002237815001415 (15pp.)
\bibitem {Webb5}
Webb, G. M. and Mace, R.L. 2015, Potential Vorticity in Magnetohydrodynamics, J. Plasma Phys., 81, Issue 1, article 905810115.
\bibitem {Sturrock}
P. A.  Sturrock, {\it Plasma Physics} (Cambridge University Press,
Cambridge, 1994)
\bibitem{Moffatt}
V. A. Vladimirov and H. K. Moffatt, J. Fluid. Mech. {\bf 283}
125-139 (1995)
\bibitem {Kats}
A. V. Kats, Los Alamos Archives physics-0212023 (2002), JETP Lett.
77, 657 (2003)
\bibitem {Kats3}
A. V. Kats and V. M. Kontorovich, Low Temp. Phys. 23, 89 (1997)
\bibitem {Kats4}
A. V. Kats, Physica D 152-153, 459 (2001)
\bibitem {Sakurai}
T. Sakurai,  Pub. Ast. Soc. Japan {\bf 31} 209 (1979)
\bibitem{Morrison}
P.J. Morrison, Poisson Brackets for Fluids and Plasmas, AIP Conference proceedings, Vol. 88, Table 2.
\bibitem {Yang}
W. H. Yang, P. A. Sturrock and S. Antiochos, Ap. J., {\bf 309} 383 (1986)
\bibitem {Yah}
Yahalom A., "A Four Function Variational Principle for Barotropic Magnetohydrodynamics"
EPL 89 (2010) 34005, doi: 10.1209/0295-5075/89/34005 [Los - Alamos Archives - arXiv: 0811.2309]
\bibitem {Yah2}
Asher Yahalom "Aharonov - Bohm Effects in Magnetohydrodynamics" Physics Letters A.
Volume 377, Issues 31-33, 30 October 2013, Pages 1898-1904.
\bibitem{Yahnbmhd}
Yahalom A., "Simplified Variational Principles for non-Barotropic Magnetohydrodynamics". J. Plasma Phys. (2016), vol. 82, 905820204. doi:10.1017/S0022377816000222.
\bibitem {Yahalom2}
 Asher Yahalom "Non-Barotropic Magnetohydrodynamics as a Five Function Field Theory". International Journal of Geometric Meth\-ods in Modern Physics, No. 10 (Nove\-mber 2016). Vol. 13 1650130 \copyright \ World Scientific Publishing Company, DOI: 10.1142/S0219887816501309.
  \bibitem {Yahalom3}
 Asher Yahalom "Simplified Variational Principles for Stationary non-Barotropic Magnetohydrodynamics" International Journal of Mechanics, Volume 10, 2016, p. 336-341. ISSN: 1998-4448.
  \bibitem {Yahalom4}
 A. Yahalom "Variational Principles for Non-Barotropic Magnetohydrodynamics a Tool for Evaluation of Plasma Processes" Proceedings of the XV Israeli-Russian Bi-National Workshop "The optimization of composition, structure and properties of metals, oxides, composites, nano - and amorphous materials", page 149-165, 26 - 30 September, 2016, Yekaterinburg, Russian Federation.
\bibitem {FLS}
A. Frenkel, E. Levich and L. Stilman Phys. Lett. A {\bf 88}, p.
461 (1982)
\bibitem {Zakharov}
V. E. Zakharov and E. A. Kuznetsov, Usp. Fiz. Nauk 40, 1087 (1997)
\bibitem{Bekenstien}
J. D. Bekenstein and A. Oron, Physical Review E Volume 62, Number 4, 5594-5602 (2000)
\bibitem {Kats2}
A. V. Kats, Phys. Rev E 69, 046303 (2004)
\bibitem{YahPinhasKop}
A. Yahalom, G. A. Pinhasi and M. Kopylenko, "A Numerical Model Based on Variational Principle for Airfoil and Wing Aerodynamics",
proceedings of the AIAA Conference, Reno, USA (2005).
\bibitem {Bateman}
 H. Bateman "On Dissipative Systems and Related Variational Principles" Phys. Rev. 38, 815 – Published 15 August 1931.
 \bibitem{Clebsch1}
Clebsch, A., Uber eine allgemeine Transformation der hydro-dynamischen Gleichungen.
{\itshape J.~reine angew.~Math.}~1857, {\bf 54}, 293--312.
\bibitem{Clebsch2}
Clebsch, A., Uber die Integration der hydrodynamischen Gleichungen.
{\itshape J.~reine angew.~Math.}~1859, {\bf 56}, 1--10.
\bibitem{Katsb}
A.V. Kats,"Canonical description of ideal magnetohydrodynamic flows and integrals of motion" PRE 69, 046303 (2004).
\bibitem {AhBo}
Y. Aharonov and D. Bohm "Significance of Electromagnetic Potentials in the Quantum Theory"
Physical Review Vol. 115 No. 3 (1959) pages 485-491.
\bibitem {Oudenaarden}
Alexander van Oudenaarden, Michel H. Devoret, Yu. V. Nazarov \& J. E. Mooij
"Magneto-electric Aharonov–Bohm effect in metal rings"  NATURE, 391, p. 768-770 (1998).
\bibitem {Akira}
Akira Tonomura, Tsuyoshi Matsuda, Byo Suzuki, Akira Fukuhara,\hfill \\ Nobuyuki Osakabe,
Hiroshi Umezaki, Junji Endo, Kohsei Shinagawa, Yu\-taka Sugita, and Hideo Fujiwara
"Observation of Aharonov-Bohm Effect by Electron Holography" PRL, 48, 21, p. l443-1446, (1982).
\bibitem {Berry}
M.V. Berry, R.G. Chambers, M. D. Large, C. Upstill and J. C. Walmsley "Wavefront
dislocations in the Aharonov-Bohm effect and its water wave analogue", Eur. J. Phys. 1, 154-162, (1980).
\bibitem{Woltjer}
Woltjer L. "Hydromagnetic Equilibrium. IV. Axisymmetric Compressible Media" ApJ, vol. 130, p. 405, (1959).
\bibitem {Boozer}
Boozer A. H.,"Physics of magnetically confined plasmas" {\it Rev. Mod. Phys.}, vol 76, 1071, 2004.
\bibitem {18}
Bohm D. Physical Review Vol. 85, 166-179, 1952.
\bibitem {19}
Bohm D. Physical Review Vol. 85, 180-193, 1952.
\bibitem {YahAB}
Yahalom A., "Barotropic Magnetohydrodynamics as a Four Function Field Theory with Non-Trivial Topology and Aharonov-Bohm Effects"
Proceedings of the Sixth International Conference on Mathematical
Modelling and Computer Simulation of Materials Technologies MMT 2010, Part 1 287-296, Ariel, Israel. [arXiv:1005.3977].
\bibitem{VM}
Vladimirov V. A.  and Moffatt  H. K. , J. Fluid. Mech. {\bf 283} 125-139 (1995)
\bibitem {Moffatt78}
Moffatt H. K., \textit{Magnetic field generation in electrically conducting fluids},
Monograph in Cambridge University Press series on Mechanics and Applied Mathematics, 1978.
\bibitem {ArKh}
Arnold V. I. \& Khesin B. A., \textit{Topological Methods in Hydrodynamics},p. 177, Springer,  1998.
\bibitem {Taylor}
Taylor J. B. "Relaxation of Toroidal Plasma and Generation of Reverse Magnetic Fields,"
Physical Review Letters, Volume 33, Number 19,  4 November 1974.
\bibitem{Biskamp}
Biskamp D., \textit{Nonlinear Magnetohydrodynamics}, Cambridge Monographs on Plasma Physics(No. 1), July 1997.
\bibitem{DuMa}
Duez V. \& Mathis S., "Relaxed equilibrium configurations to model fossil fields I. A first family,"
Astronomy \& Astrophysics, vol. 517, id. A58, 2010.
\bibitem {Braithwaite}
Braithwaite J., "On non-axisymmetric magnetic equilibria in stars," MNRAS, vol. 386, 1947-1958, 2008.
\bibitem {Pipin}
Pipin V. V.  and Kosovichev A. G. "The Asymmetry of Sunspot Cycles and Waldmeier Relations as
a Result of Nonlinear Surface-Shear Shaped Dynamo," The Astrophysical Journal, 741:1 (9pp), 2011.
\bibitem {SolPhys}
Rodiger G., Kitchatinov L.L. \&  Brandenburg A., "Cross Helicity and Turbulent Magnetic Diffusivity
in the Solar Convection Zone,"  Solar Physics, 269, 3-12 (2011), DOI 10.1007/s11207-010-9683-4.
\bibitem {Wedemeyer}
Wedemeyer-Bohm, S., Skullion, E., Steiner, O., Rouppe van der Voort, L., la Cruz Rodriguez, J., Fedun, V., and Erdelyi, R. 2012, Magnetic tornadoes as energy channels into the solar corona, Nature, 486, June 28 issue, pp 505-508.

\end {thebibliography}
\end {document}